# Nontrivial impact of interlayer coupling on thermal conductivity: opposing trends in in-plane and out-of-plane phonons

H. F. Feng,[1] B. Liu,[1] J. L. Bai,[1] X, Zhang,[2] Z. X. Song,[1] and Zhi-Xin Guo[1,*]

[1]*State Key Laboratory for Mechanical Behavior of Materials, School of Materials Science and Engineering, Xi'an Jiaotong University, Xi'an, Shanxi, 710049, China.*
[2]*College of Mechanical and Materials Engineering, Xi'an University, Xi'an 710065, China.*

[*]zxguo08@xjtu.edu.cn

**Abstract**

 The study of heat transport in two-dimensional (2D) materials reveals novel behaviors due to quantum confinement effects, where in-plane and out-of-plane phonons play crucial roles. In 2D materials like graphene, it is widely recognized that the out-of-plane vibrational mode is the primary contributor to thermal conductivity owing to the mirror symmetry. Based on this perspective, the introduction of interlayer coupling, which breaks this symmetry, is expected to induce a significant reduction in thermal conductivity within 2D materials. Nevertheless, recent studies have presented unexpected findings, indicating that interlayer coupling can actually increase thermal conductivity of 2D materials. This controversial result suggests a nontrivial underlying mechanism governing the effects of interlayer coupling on thermal conductivity in 2D materials, necessitating further exploration. In our work, we investigate the modulation of thermal conductivity through interlayer coupling in a sandwich structure composed of hexagonal boron nitride (h-BN) and bilayer graphene (BG), specifically a h-BN/BG/h-BN system. Through molecular dynamics simulations, we find that the thermal conductivity from out-of-plane phonons can be significantly reduced, while that from in-plane phonons can be significantly increased, as the interlayer coupling strength increases. This results in a nontrivial, coupling-strength-dependent overall thermal conductivity. The phonon spectrum analysis conducted using our modified package reveals that the upshift and flattening of the out-of-plane (ZA and ZO) phonon modes are mainly responsible for these variations, and the extent of the upshift and flattening is proportional to the strength of interlayer coupling. This work offers new insights into manipulating the thermal conductivity of 2D materials.

Over the years, two-dimensional (2D) materials have gained significant attention due to their unique properties and potential technological applications [1-12]. The high lattice thermal conductivity of 2D graphene and other carbon nanostructures has stimulated intensive studies to understand phonon transport in them [13-18]. Beyond the potential for thermal management in nano-electronic devices, graphene can serve as an invaluable benchmark for investigating fundamental problems of thermal transport in low-dimensional systems [19]. Extensive investigations have predicted various anomalous heat transfer phenomena within the 2D lattice model, particularly concerning the contributions of in-plane and out-of-plane phonons to thermal conductivity [19-25].

Generally, acoustic phonons are crucial to the thermal conductivity of 2D materials. The acoustic phonons can be characterized by two distinct types: those vibrating within the layer's plane, featuring transverse and longitudinal acoustic (TA and LA) phonons, and those vibrating out of the plane, known as flexural (ZA and ZO) phonons. In monolayer 2D materials like graphene, it is generally accepted that the ZA mode is the primary contributor to thermal conductivity (71-89%), owing to the mirror symmetry in the out-of-plane direction, which promotes thermal transport by forbidding certain phonon scattering channels [19-23]. Therefore, it can be inferred that the symmetry disruptions induced by interlayer coupling can significantly impact the contribution of ZA mode to the thermal conductivity.

A straightforward example is bilayer graphene (BG) with an AB stacking structure, where the mirror symmetry is broken [21]. Indeed, a significant reduction in thermal conductivity of BG compared to monolayer graphene has been found, attributed to a notable decrease in thermal conductivity arising from ZA and ZO modes [26,27]. According to this effect, with the increase of interlayer coupling, the thermal conductivity would continue to decrease. However, recent researches show that thermal conductivity can unexpectedly increase with the increase of interlayer coupling strength [28-32]. This conflicting result indicates the existence of a nontrivial mechanism for the interlayer coupling effect on the thermal conductivity of 2D material, which is in great need to be revealed.

In this study, we employ molecular dynamics simulations to investigate the effect of interlayer coupling on the thermal conductivity of BG within a hexagonal boron nitride (h-BN)/BG/h-BN sandwich structure. The interlayer coupling strength is modulated by adjusting the distance between the h-BN layers. We find that interlayer coupling significantly affects the thermal conductivity contributions from both out-of-plane ($\kappa_{out}$) and in-plane phonons ($\kappa_{in}$). Specifically, as the interlayer coupling strength increases, $\kappa_{out}$ decreases while $\kappa_{in}$ increases, leading to an extraordinarily high total thermal conductivity. Further analysis reveals that these variations primarily arise from the upshift and flattening of the ZA and ZO phonon modes, and the extents of upshift and flattening are proportional to the strength of interlayer coupling.

Our calculations for thermal conductivity are based on the homogeneous nonequilibrium molecular dynamics (HNEMD) method (see Sec. SI in the Supplemental Material [33]). We first investigate the interlayer coupling effect on thermal conductivity in freestanding BG. Fig. S1 shows the thermal conductivity of both monolayer graphene and BG in an AB stacking order. It is found that the monolayer graphene has a high thermal conductivity of 2795.1 W/mK, with a dominating contribution of 72% (1984.0 W/mK) from out-of-plane phonons (mostly attributed to the ZA mode [20]). As for the BG, thermal conductivity drops to 2268.7 W/mK, and the contribution of out-of-plane phonons is reduced to 59% (1339.3 W/mK). This result aligns with previous findings, indicating that the impact of interlayer coupling mainly lies in $\kappa_{out}$ [21,44]. However, it is noted that $\kappa_{in}$ is also affected by the interlayer coupling in BG, which has been overlooked in previous studies owing to its relatively small values. Our results show that the interlayer coupling leads to an increase in $\kappa_{in}$ by about 15% (from 811.1 W/mK for monolayer graphene to 929.4 W/mK for freestanding BG). This result demonstrates that the interlayer coupling has opposite impacts on the thermal conductivity of BG, i.e., decreasing $\kappa_{out}$ but increasing $\kappa_{in}$.

The above finding prompts us to explore how variations in interlayer coupling affect $\kappa_{in}$ and $\kappa_{out}$. To effectively modulate the interlayer coupling in BG, we employ a mechanical tuning method by utilizing the substrate to exert pressure. Our approach involves the creation of a heterostructure consisting of h-BN and BG, forming a h-BN/BG/h-BN sandwich structure [Figs. 1(a) and 1(b)]. Within this configuration, we arrange the two graphene layers in an AB stacking order, and the h-BN atomic lattice correspondingly aligns with the adjacent graphene layer. To preserve the stability of interlayer interactions throughout our simulations, we restrict the vibrations of the h-BN layers. This design allows precise control over the distance between BG ($d_{int}$) by adjusting the spacing between the h-BN layers, which in turn alters the interlayer coupling in BG.

In the above simulation process of freestanding BG, the interlayer distance is maintained at about 3.4 Å. To facilitate a direct comparison, we initially select a sandwich structure with $d_{int}$ = 3.4 Å for analysis. Notably, Fig. 1(c) reveals a decrease in $\kappa_{out}$ for this structure compared to the freestanding BG. This reduction can be attributed to the additional h-BN layer, which further disrupts the symmetry of the graphene. Unexpectedly, despite this decrease in $\kappa_{out}$, the total thermal conductivity experiences an increase due to a significant rise in $\kappa_{in}$. As shown in Fig. 1(d), it is unexpected that in-plane phonons predominate over out-of-plane phonons in their contribution to the total thermal conductivity. As we further enhance the interlayer coupling, $\kappa_{in}$ continues to increase while $\kappa_{out}$ progressively diminishes. From freestanding BG to the sandwich structure with $d_{int}$ = 3.1 Å, the growth in $\kappa_{in}$ outweighs the decline in $\kappa_{out}$, leading to an overall increase in the total thermal conductivity of BG. However, as the $d_{int}$ is reduced from 3.1 Å to 2.8 Å, the gain in $\kappa_{in}$ is less pronounced than the loss in $\kappa_{out}$, resulting in a subsequent decrease in the total thermal conductivity. Consequently, the proportion of $\kappa_{in}$ contributing to the total thermal conductivity rises substantially from 41.0% for freestanding BG to 83.6% for BG in the sandwich structure with $d_{int}$ = 2.8 Å. This pronounced disparity in $\kappa_{in}$ and $\kappa_{out}$ highlights the opposing effects of interlayer coupling on in-plane versus out-of-plane phonons.

The change in thermal conductivity can be reflected in the phonon spectrum to some extent, so we examine the phonon spectrum of both freestanding BG and BG in the sandwich structures with varying interlayer distances for comparison. It should be noted that the general phonon calculation procedures based on finite displacement theory (e.g. implemented in the Phonopy package [35]) cannot directly get the phonon spectrum of BG from the sandwich structure, so we modify the calculation process of this package to make it possible (see Sec. SI in the Supplemental Material). Fig. 2(a) illustrates the phonon spectrum of freestanding BG, where the ZA mode splits at the Γ point, giving rise to the appearance of the ZO mode, which aligns with prior findings [27]. Upon the addition of the h-BN substrate ($d_{int}$ = 3.4 Å), ZA and ZO modes experience upshift and flattening near the Γ point, while the remaining acoustic and optical modes remain largely unchanged [Fig. 2(b)]. Similarly, as the interlayer distance decreases ($d_{int}$ = 3.1 Å and 2.8 Å), the ZA and ZO modes experience a further upshift and flattening, while the other modes generally stay constant except for slight splitting around the Γ point [Figs. 2(c) and 2(d)]. The divergent responses of in-plane and out-of-plane (ZA/ZO) modes provide compelling evidence for the contrasting trends observed in $\kappa_{in}$ and $\kappa_{out}$. However, the phonon spectrum can only provide a qualitative understanding. Note that the thermal conductivity is proportional to the product of the number of excited phonons (NEP), phonon group velocity, and phonon mean free path (MFP), all of which are phonon frequency ($\omega$) dependent [29,45]. For a more comprehensive analysis, we systematically examine these factors that influence thermal conductivity. Such step-by-step verification can move beyond qualitative insights to more definitive evidence supporting our observations.

We begin our analysis by examining NEP. As shown in Fig. 3(a), the NEP attributed to in-plane phonons hardly changes with interlayer coupling strength, indicating that the NEP is not driving $\kappa_{in}$ changes. Conversely, Fig. 3(b) demonstrates that as interlayer coupling strengthens, the NEP of out-of-plane phonons not only shifts to higher frequencies but also diminishes significantly in amplitude. We further integrate the NEP (out-of-plane phonons) across the phonon frequency range as the total count of NEP, the integration results for the sandwich structures with $d_{int}$ decreasing from 3.4 Å to 2.8 Å correspond to 65.2%, 50.9%, and 35.6% of that for freestanding BG, respectively. The reduction in NEP provides supportive evidence for the decrease in $\kappa_{out}$ from dynamic calculations.

Phonon group velocity is another crucial factor influencing thermal conductivity. In our study, the group velocity of in-plane phonons displays minimal variation across different structures, from freestanding BG to sandwich structures with different $d_{int}$ [Fig. 3(c)]. This observation aligns with the robust phonon spectrum of in-plane vibrations, suggesting that changes in the group velocity of in-plane phonons do not significantly influence the $\kappa_{in}$. Contrarily, Fig. 3(d) reveals that the group velocity of out-of-plane phonons has an overall movement towards higher frequency ranges, attributed to the upshift of ZA and ZO modes. Meanwhile, the group velocity of out-of-plane phonons is suppressed below 25 THz owing to the flattening of ZA and ZO modes. Compared to the group velocity of ZA and ZO modes in freestanding BG, the maximum value decreases by 2.5%, 7.6%, and 18.5%, and the average value fells by 2.2%, 18.4%, and 42.6%, corresponding to sandwich structure with $d_{int}$ = 3.4 Å, 3.1 Å and 2.8 Å, respectively. These results indicate that the addition of a substrate has little impact on the group velocity of out-of-plane phonons, whereas the enhancement of interlayer coupling significantly reduces it. Hence, group velocity plays a minor role in $\kappa_{out}$ shifts in cases of weak interlayer coupling, while its decrease plays a relatively significant role in the reduction of $\kappa_{out}$ under strong interlayer coupling conditions.

From the above discussions, it is evident that the influence of interlayer coupling induced by substrates on NEP and phonon group velocity is significant for $\kappa_{out}$ but insignificant for $\kappa_{in}$. Then we explore the impact of interlayer coupling on phonon MFP, another crucial factor for thermal conductivity. Figs. 4(a) and 4(b) show the calculated MFP of the BG for both in-plane and out-of-plane phonons. At low frequencies, the MFP of in-plane phonons increases notably with the addition of substrates [Fig. 4(a)], emerging as the predominant reason for the increase of $\kappa_{in}$. On the contrary, the MFP of out-of-plane phonons experiences a marked reduction when interlayer coupling is increased [Fig. 4(b)], serving as an additional factor for the reduction of $\kappa_{out}$. However, the changes in MFP cannot be directly obtained from conventional cognition. In the following, we delve into the change in MFP and its influence on $\kappa_{in}$ and $\kappa_{out}$.

In order to provide a clearer way to present the effects of multiple factors on thermal conductivity, we conduct a spectral decomposition analysis for both $\kappa_{in}$ and $\kappa_{out}$ [42,43]. Fig. 4(c) illustrates $\kappa_{in}$ as a function of frequency. Compared with freestanding BG, the addition of a substrate leads to a noticeable enhancement in $\kappa_{in}$ at lower frequencies, across all levels of interlayer coupling. Furthermore, the enhancement of interlayer coupling extends the frequency range over which thermal conductivity contributions significantly increase.

To clarify the effects of substrate addition on $\kappa_{in}$, we begin by comparing the BG with the $d_{int}$ = 3.4 Å sandwich structure [Fig. 4(c)], it is found that $\kappa_{in}$ increases slightly over a large frequency range and significantly below 10 THz. This suggests that the substrate predominantly increases $\kappa_{in}$ within this low-frequency range, which can be interpreted through the phonon spectrum and scattering mechanisms. It is important to recognize that out-of-plane phonons (ZA/ZO modes) are also involved in the scattering of in-plane phonons (TA/LA modes). Specifically, in BG, a significant kind of phonon scattering channel is ZA/ZO + ZA/ZO ↔ TA/LA [21]. The overlay of two key elements related to these channels is responsible for the reduction of the scattering rate, ultimately resulting in an increase of BG in $\kappa_{in}$ from freestanding BG to sandwich structure with $d_{int}$ = 3.4 Å: (1) The reduction in the NEP of out-of-plane phonons [Fig. 3(b)] leads to fewer phonons available for these scattering channels, naturally diminishing in-plane phonon scattering across the entire frequency spectrum, which explains the overall rise in MFP and $\kappa_{in}$. (2) Although the in-plane modes in the phonon spectrum remain mostly unaltered, the out-of-plane ZA and ZO modes lift to frequencies of 2.9 THz and 4.5 THz, respectively [Fig. 2(b)]. According to energy conservation laws, only in-plane modes above 5.8 THz and 9.0 THz can engage with ZA + ZA ↔ TA/LA and ZO + ZO ↔ TA/LA scattering processes. Consequently, the possibility of in-plane phonon scattering decreases significantly within the low-frequency range, leading to an increase of in-plane phonon lifetimes. Since the in-plane phonon group velocity exhibits negligible change [Fig. 3(c)], the phonon MFP (product of group velocity and phonon lifetimes) of BG in the sandwich structure significantly increases compared to the freestanding BG [Fig. 4(a)]. This discussion on phonon scattering below 9.0 THz accounts for the obvious enhancement of $\kappa_{in}$ in the 0-10

THz [Fig. 4(c)], and the frequency range is almost the same. These two mechanisms clearly explain the overall increase in $\kappa_{in}$ and the particularly significant enhancement at low frequencies.

The continued increase in $\kappa_{in}$ with enhanced interlayer coupling can also be attributed to the combined effect of the two mechanisms. As the $d_{int}$ decreases from 3.4 Å to 2.8 Å, on one hand, there is a corresponding decrease in NEP of out-of-plane phonons. This leads to a reduction in the number of phonons participating in the in-plane scattering, so that MFP and $\kappa_{in}$ continue to rise in the whole frequency range. On the other hand, the continuous upshift of the ZA and ZO modes expands the frequency range of significant thermal conductivity improvement from 0-10 THz to 0-25 THz and eventually to 0-40 THz. Such enhancements are about twice the frequency of the ZO mode at the Γ point, further validating the rationality of our phonon scattering analysis related to the changes in thermal conductivity. It is important to note that we observe a decrease of $\kappa_{in}$ in the low-frequency region for sandwich structure with 2.8 Å. This decrease is related to the splitting of the LA and TA modes induced by the strong interlayer coupling [Fig. S2], which enhances phonon scattering in the low-frequency region and reduces thermal conductivity. Nonetheless, this reduction effect is relatively minor, and the overall trend of $\kappa_{in}$ continues to increase compared to the case of $d_{int}$ = 3.1 Å.

After elucidating the increasing trend of $\kappa_{in}$, we now focus on the decline of $\kappa_{out}$ from the perspective of phonon scattering and phonon MFP. As shown in Fig. 4(d), it is noted that as interlayer coupling intensifies, the spectral contribution to $\kappa_{out}$ shifts towards higher frequencies with a notable decrease in magnitude in the 0-20 THz range. These observed variations in $\kappa_{out}$ are consistent with our previous discussions on NEP, group velocities, and MFP, but the significant reduction in MFP of out-of-plane phonons requires further investigation to fully understand its impact. As shown in Fig. 2(b), the presence of substrates flattens the ZA and ZO modes near the Γ point, facilitating resonant four-phonon scattering of the ZA + ZO ↔ ZA + ZO channel [6]. This leads to pronounced four-phonon scattering, shortening MFP of out-of-plane phonons and thus reducing $\kappa_{out}$. Additionally, as interlayer coupling grows stronger, the prevalence of this flattening spreads in reciprocal space [Figs. 2(c) and 2(d)], creating a broader zone for resonant four-phonon scattering, leading to a more pronounced reduction in the MFP of out-of-plane phonons. As a result, the interlayer coupling leads to a significant reduction of MFP of out-of-plane phonons and thus the reduction of $\kappa_{out}$.

In conclusion, our study investigates the underlying mechanisms of the influence of interlayer coupling on the thermal transport properties for in-plane and out-of-plane phonons. Based on HNEMD simulations for freestanding BG and h-BN/BG/h-BN sandwich structures, we find that increased interlayer coupling diminishes $\kappa_{out}$ but increases $\kappa_{in}$. As a result, the total thermal conductivity can be unexpectedly elevated with the presence of interlayer coupling. For the thermal transport of in-plane phonons, the enhancement of $\kappa_{in}$ mainly originates from the extended phonon MFP. On the contrary, the reduction of $\kappa_{out}$ results from three combined effects: reduction in NEP, decrease in group velocity, and decline in MFP. Phonon spectrum analysis further reveals that these variations can be attributed to the upshift and flattening of the ZA and ZO phonon modes, wherein the degree of upshift and flattening are proportional to the increase in interlayer coupling strength. Finally, we would like to remark that the observed phenomenon of the contrasting response of $\kappa_{in}$ and $\kappa_{out}$ to the interlayer coupling in BG is also applicable to other 2D materials, particularly those where out-of-plane phonons dominate the thermal transport.

We acknowledge financial support from the Natural Science Foundation of China (Grant No. 12074301), the Natural Science Foundation of Shaanxi Province (Grant No. 2023-JC-QN-0768), and Science Fund for Distinguished Young Scholars of Shaanxi Province (No. 2024JC-JCQN-09).

## References

[1] K. S. Novoselov, A. Mishchenko, A. Carvalho, and A. H. Castro Neto, Science **353**, aac9439 (2016).
[2] J. R. Schaibley, H. Yu, G. Clark, P. Rivera, J. S. Ross, K. L. Seyler, W. Yao, and X. Xu, Nat. Rev.

Mater. **1**, 16055 (2016).

[3] P. Li, J. Cao, and Z.-X. Guo, J. Mater. Chem. C **4**, 1736 (2016).

[4] M. Gibertini, M. Koperski, A. F. Morpurgo, and K. S. Novoselov, Nat. Nanotechnol. **14**, 408 (2019).

[5] M. Long, P. Wang, H. Fang, and W. Hu, Adv. Funct. Mater. **29**, 1803807 (2019).

[6] L. Xie, J. H. Feng, R. Li, and J. Q. He, Phys. Rev. Lett. **125**, 245901 (2020).

[7] J. Huang, P. Li, X. Ren, and Z.-X. Guo, Phys. Rev. Appl. **16**, 044022 (2021).

[8] H. Zhang, Y. Wang, W. Yang, J. Zhang, X. Xu, and F. Liu, Nano Lett. **21**, 5828 (2021).

[9] B. Liu, X. X. Ren, X. Zhang, P. Li, Y. Dong, and Z.-X. Guo, Appl. Phys. Lett. **122**, 152408 (2023).

[10] P. Li, B. Liu, S. Chen, W.-X. Zhang, and Z.-X. Guo, Chin. Phys. B **33**, 017505 (2024).

[11] S. Yan, K. Wang, Z. Guo, Y.-N. Wu, and S. Chen, Nano Lett. **24**, 6158 (2024).

[12] X. Zhang, B. Liu, J. Huang, X. Cao, Y. Zhang, and Z.-X. Guo, Phys. Rev. B **109**, 205105 (2024).

[13] A. A. Balandin, S. Ghosh, W. Bao, I. Calizo, D. Teweldebrhan, F. Miao, and C. N. Lau, Nano Lett. **8**, 902 (2008).

[14] S. Ghosh, I. Calizo, D. Teweldebrhan, E. P. Pokatilov, D. L. Nika, A. A. Balandin, W. Bao, F. Miao, and C. N. Lau, Appl. Phys. Lett. **92**, 151911 (2008).

[15] Z. Guo, D. Zhang, and X.-G. Gong, Appl. Phys. Lett. **95**, 163103 (2009).

[16] Z.-x. Guo and X.-g. Gong, Front. Phys. **4**, 389 (2009).

[17] A. A. Balandin, Nat. Mater. **10**, 569 (2011).

[18] A. M. Marconnet, M. A. Panzer, and K. E. Goodson, Rev. Mod. Phys. **85**, 1295 (2013).

[19] Z. Fan, LuizFelipeC. Pereira, P. Hirvonen, M. M. Ervasti, K. R. Elder, D. Donadio, T. Ala-Nissila, and A. Harju, Phys. Rev. B **95**, 144309 (2017).

[20] L. Lindsay, D. A. Broido, and N. Mingo, Phys. Rev. B **82**, 115427 (2010).

[21] L. Lindsay, D. A. Broido, and N. Mingo, Phys. Rev. B **83**, 235428 (2011).

[22] Z. Wei, J. Yang, K. Bi, and Y. Chen, J. Appl. Phys. **116**, 153503 (2014).

[23] M. Gill-Comeau and L. J. Lewis, Phys. Rev. B **92**, 195404 (2015).

[24] X. Gu, Y. Wei, X. Yin, B. Li, and R. Yang, Rev. Mod. Phys. **90**, 041002 (2018).

[25] LuizFelipeC. Pereira and D. Donadio, Phys. Rev. B **87**, 125424 (2013).

[26] K. Sato, J. S. Park, R. Saito, C. Cong, T. Yu, C. H. Lui, T. F. Heinz, G. Dresselhaus, and M. S. Dresselhaus, Phys. Rev. B **84**, 035419 (2011).

[27] B. D. Kong, S. Paul, M. B. Nardelli, and K. W. Kim, Phys. Rev. B **80**, 033406 (2009).

[28] H. F. Feng, B. Liu, and Z.-X. Guo, Phys. Rev. B **108**, L241405 (2023).

[29] Z.-X. Guo, D. Zhang, and X.-G. Gong, Phys. Rev. B **84**, 075470 (2011).

[30] R. Su, Z. Yuan, J. Wang, and Z. Zheng, Phys. Rev. E **91**, 012136 (2015).

[31] S. S. PaulChowdhury, A. Samudrala, and S. Mogurampelly, Phys. Rev. B **108**, 155436 (2023).

[32] Z.-Y. Ong and E. Pop, Phys. Rev. B **84**, 075471 (2011).

[33] See Supplemental Material at http://link.aps.org/supplemental/xx.xxxxxx for the computational method; thermal conductivity of monolayer graphene and BG; phonon spectrum of BG for sandwich structure with $d_{int}$ = 2.8 Å near Γ point. The Supplemental Material also contains Refs. [34-43].

[34] K. Momma and F. Izumi, J. Appl. Crystallogr. **44**, 1272 (2011).

[35] A. Togo, L. Chaput, T. Tadano, and I. Tanaka, J. Phys.: Condens. Matter **35**, 353001 (2023).

[36] L. Lindsay and D. A. Broido, Phys. Rev. B **81**, 205441 (2010).

[37] C. Si, X.-D. Wang, Z. Fan, Z.-H. Feng, and B.-Y. Cao, Int. J. Heat Mass Transfer **107**, 450 (2017).
[38] L. A. Girifalco, M. Hodak, and R. S. Lee, Phys. Rev. B **62**, 13104 (2000).
[39] C.-Y. Chen, Y. She, H. Xiao, J. Ding, J. Cao, and Z.-X. Guo, J. Phys.: Condens. Matter **28**, 145003 (2016).
[40] X. Gu, Z. Fan, H. Bao, and C. Y. Zhao, Phys. Rev. B **100**, 064306 (2019).
[41] Z. Fan, L. F. C. Pereira, H.-Q. Wang, J.-C. Zheng, D. Donadio, and A. Harju, Phys. Rev. B **92**, 094301 (2015).
[42] Z. Fan, H. Dong, A. Harju, and T. Ala-Nissila, Phys. Rev. B **99**, 064308 (2019).
[43] A. J. Gabourie, Z. Fan, T. Ala-Nissila, and E. Pop, Phys. Rev. B **103**, 205421 (2021).
[44] D. Singh, J. Y. Murthy, and T. S. Fisher, J. Appl. Phys. **110**, 044317 (2011).
[45] S. G. Wang, H. F. Feng, and Z.-X. Guo, Comput. Mater. Sci **228**, 112345 (2023).

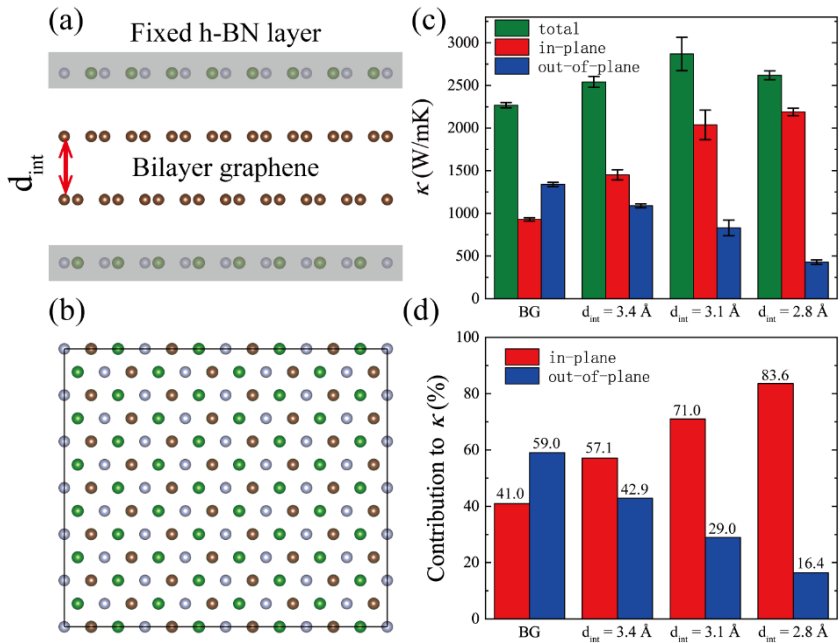

Fig. 1. Schematic structures of BG sandwiched between two fixed h-BN layers for (a) side view and (b) top view. $d_{int}$ represents the interlayer distance between two graphene layers. The Green, saddle brown, and gray balls represent the B, C, and N atoms, respectively. The fixed h-BN layers are indicated by the semitransparent rectangle. (c) Total thermal conductivity $\kappa$, thermal conductivity from in-plane and out-of-plane phonons for freestanding BG and BG in sandwich structures with different $d_{int}$. (d) Thermal conductivity contribution of in-plane and out-of-plane phonons for freestanding BG and BG in sandwich structures with different $d_{int}$.

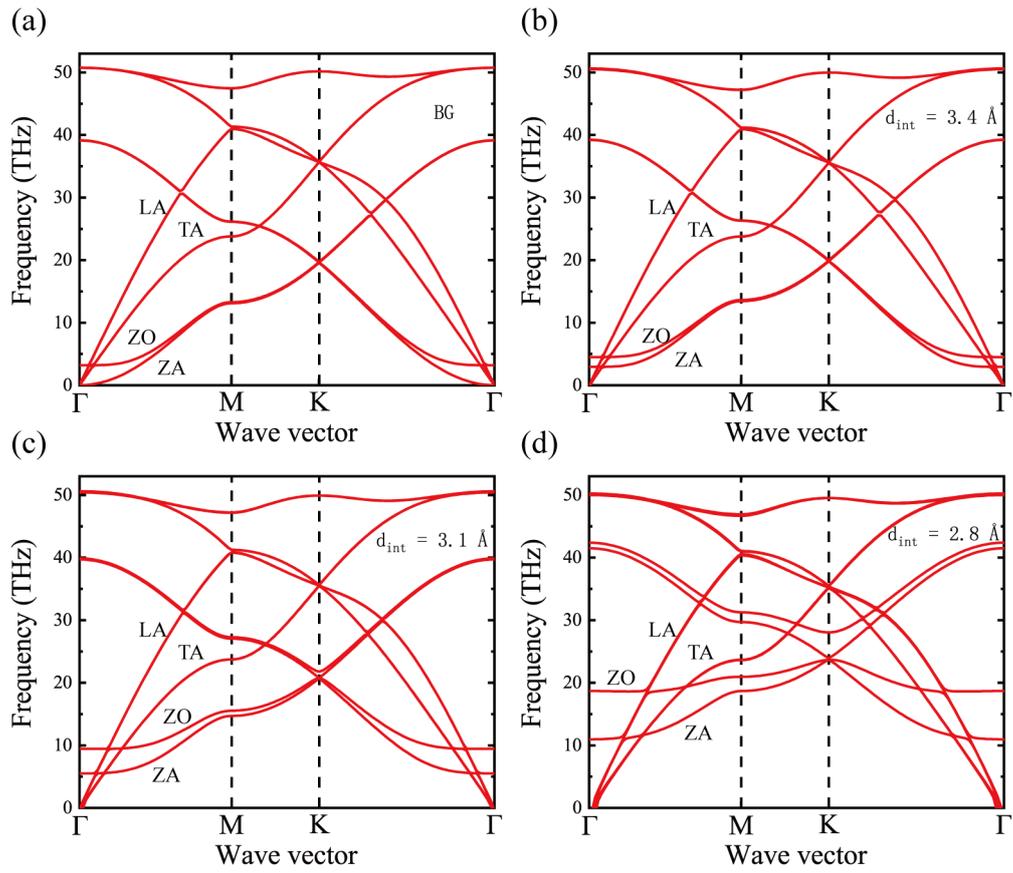

Fig. 2. Phonon spectrum of (a) freestanding BG and BG in sandwich structures with (b) $d_{int}$ = 3.4 Å, (c) $d_{int}$ = 3.1 Å and (d) $d_{int}$ = 2.8 Å.

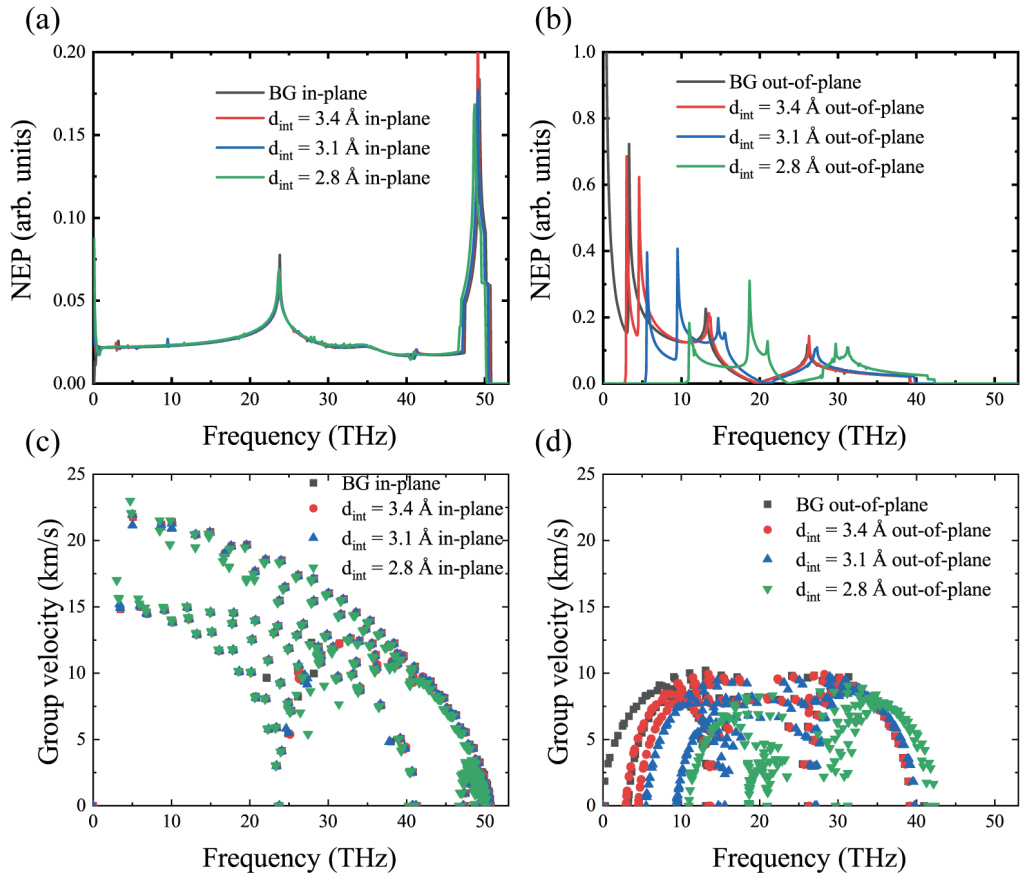

Fig. 3. NEP of freestanding BG and BG in sandwich structures with different $d_{int}$ for (a) in-plane and (b) out-of-plane phonons. Group velocity of freestanding BG and BG in sandwich structures with different $d_{int}$ for (c) in-plane and (d) out-of-plane phonons.

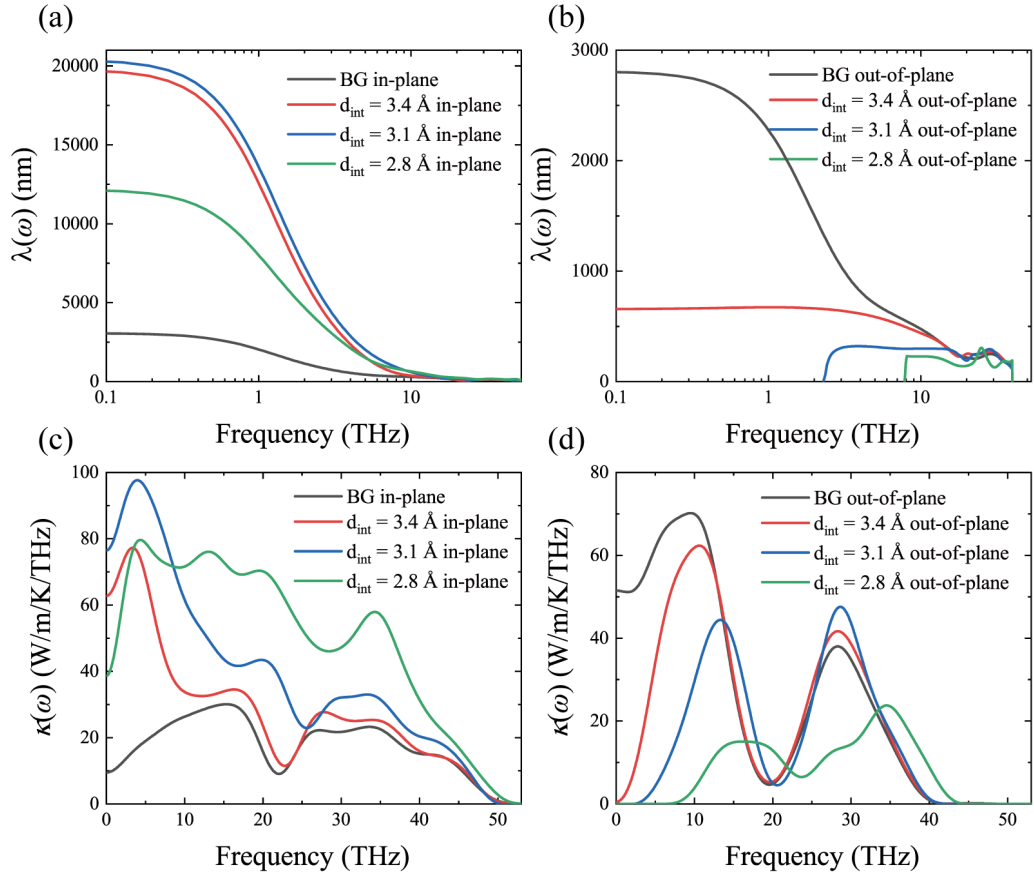

Fig. 4. The Spectral phonon mean free path (MFP) $\lambda(\omega)$ of freestanding BG and BG in sandwich structures with different $d_{int}$ for (a) in-plane and (b) out-of-plane phonons. The spectral thermal conductivity $\kappa(\omega)$ of freestanding BG and BG in sandwich structures with different $d_{int}$ for (c) in-plane and (d) out-of-plane phonons.